# ANALYSES OF KLYSTRON MODULATOR APPROACHES FOR NLC [*]

Anatoly Krasnykh, SLAC, Box 4349 Stanford, CA 94309

*Abstract*

Major changes to the Next Linear Collider (NLC) design were facilitated by the experimental testing of the 75 MW X-band klystron at a 3.0 μsec pulse width [1] and new component development allowing the delay line distribution system (DLDS) to operate with eight bins instead of four. This change has a direct effect on the design of the klystron modulator. The general approaches, which are being studied intensively, are: the conventional base line modulator with two klystrons [2], a Hybrid version of the baseline with a solid-state on/off switch, a solid-state induction type modulator that drives eight klystrons [3], and a solid-state direct switch modulator [4]. Some form of pulse transformer is the matching element between the klystron beam and the energy store in the all of these approaches except the direct switch. The volume and cost of the transformer is proportional to the peak pulse power and the output pulse width. The recent change in the NLC design requires double the transformer effective core area, and increase both the size and cost of modulator. In the direct switch model there is no pulse transformer. The klystron beam potential is practically equal to the potential of the energy storage element. Here the solid-state switch blocks the 500 kV DC voltage of the storage element. In this paper transformerless modulator approaches are presented based upon a Marx method of voltage multiplication using on/off Insulated Gate Bipolar Transistors (IGBT's) instead of on switches. DC voltage power supply system is much simpler as compared to the power system of the direct switch approach.

## INTRODUCTION

The current design of the future linear colliders consists of two X-band linacs powered by over 1,600 klystrons. The facility will consume in excess of 200 MW of average power. The total number of NLC klystrons and modulators can be reduced if RF pulse duration is longer. The pulse width for NLC klystrons is not precisely defined for the time being. The present klystron modulator for NLC requires a 500 kV, 530 A, 3 microsecond flat top pulse with 120 PPS to drive a pair of 75 MW klystrons. These parameters are used in planning for the NLC, but R&D is continuing on klystron improvements. The design of multiple beam klystrons with lower beam potential is an example of a possible improvement. The modulator for such a klystron or group of klystrons will be simpler, reliable, and less expensive.

## ENERGY TRANSFER EFFICIENCY

The major design requirements for the modulator are cost, efficiency, reliability. The efficiency is one important parameter due to large number of modulators for NLC. It is useful to identify losses into two separate efficiencies. The first component is a ratio of the energy delivered into klystron beam and PFN storage energy. It is the discharge energy efficiency. A second component is a ratio of the flat top portion of the klystron pulse energy and the energy of pulse width. It is energy transfer efficiency. The efficiency of modulator is a product of these components.

It was shown [2] that the discharge energy efficiency of ~92-93% can be obtained for PFN using oil filled capacitors, two gap thyratron and conventional pulse transformer for the 150 MW and 1.5 μsec output levels. The energy transfer efficiency is still close to ~70%. The pulse transformer is one element of modulator, which limits the energy transfer efficiency due to a limitation of the transmission band of the pulse power. It is well known that the rise time of pulse transformer is limited by the period which is determined by the leakage inductance $L$ and the distributed capacitance $C$, as following $t_1 \sim \sqrt{LC}$. The geometry of high voltage pulse transformer fixes $L$ and $C$. It is evident also that a transformer with a small number of turns ought to have a wider bandwidth. On the other hand the reduction of the number of turns produces larger pulse droop and core size for the given pulse width. Moreover, for transformers with a low $L$ and $C$ the inductance of the current loop between the PFN and a primary winding through switch $(L_c)$ and the load capacitance $(C_g)$ can play a noticeable role on the rise/fall time as well as energy transfer efficiency. For optimal case it has been found that the energy transfer efficiency can be evaluated by

$$\eta_{te} \cong \frac{1}{1+\dfrac{2t_1}{t_{rf}}} \cong \frac{1}{1+\dfrac{2\pi}{\sqrt{3}}\cdot\dfrac{w_1 p}{\lambda}\cdot\sqrt{1+\dfrac{4d}{p}}\cdot\sqrt{1+k_1}\cdot\sqrt{1+k_2}}$$

where $t_{rf}$ is the flat-top duration, $w_1$ and $p$ are the number turns and the circumference of primary winding, $d$ is the average distance between primary and secondary windings. Coefficients $k_1=L_c/L$ and $k_2=C_g/C$ are the additional parts of the primary inductance and the secondary load capacitance, which have been normalized by $L$ and $C$ of the transformer. The effective length of rf-pulse is $\lambda=ct_{rf}/(n-1)\varepsilon^{0.5}$, where $c$ is the velocity of light, $n$ is the turn ratio, and $\varepsilon$ is the dielectric constant of the media between windings. The coefficient $w_1p/\lambda<1$ is responsible for the transmission of the energy and has to be minimized (i.e. to use a media with lower $\varepsilon$, and lower turn ratio) for the given amount of $w_1p$. For example, this coefficient is ~0.035 for the conventional pulse transformer. The coefficient $(1+4d/p)^{0.5}$ is responsible for the degree of the high voltage isolation between windings and it is ~1.24. The contribution of $k_2$ is higher as compared to $k_1$ for the high power modulators.

---

[*] Work supported by Department of Energy contract DE-AC03-76SF00515

High core dynamic capacity effect can take noticeable place during transmission of the high pulse power through transformer. Actually equations, which describe the processes in the pulse transformer, assume that a velocity of field propagation in the core material is instantaneous. However the core magnetizing velocity is a function of conductivity of the ribbon, effective pulse permeability, etc. The step of the power starts to propagate in the core material from periphery to the center. The velocity of the magnetization has to be extremely fast in the beginning of process. This can happen when an extremely large spike of the magnetizing current amplitude takes place. The quality of the core material, ribbon thickness, and the core geometry are responsible for this effect. The charge of the load capacity, of the transformer distributed capacity, and of the effective core dynamic capacity goes practically simultaneously. The value of this capacity depends on $dB/dt$ and amplitude of $H(t)$ for the simple case.

In spite of described effects the demonstrated efficiency of modulator is rather high. For instance, the energy transfer is ~70% for 1.5 microsecond rf-pulse duration [2]. It is expected that the total efficiency can be improved on ~12% for 3 microsecond pulse width in the conventional modulator configuration.

The use of features of coupled transmission lines lies in the base of an idea how the energy transfer efficiency can be improved. The first approach is related to the use of coupled transmission lines, which support true TEM modes. The approach consists of a bunch of small transformers, which are connected as shown in Fig. 1a).

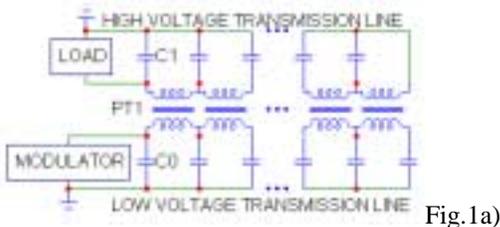
Fig.1a)

Both lines are uniform and linear for a simple case. The power step of the modulator is launched at the input of the coupled line. The number of cells and their parameters will determinate the transmission band of the pulse transformation. If the group velocity of the exciting TEM wave of the first line is equal to the group velocity of the second line, the energy of the pulse will be launched on the load with high efficiency.

Two coupled pieces of transmission lines are added to the primary and secondary winding of $PT_0$ as shown in Fig. 1b). The initial part of the pulse power is launched into the load though the fast coupling lines. The coupling is due to fast $PT_1$. The rest of power is transmitted through $PT_0$. Modulators based on schemes of Fig. 1a) and 1b) consist of cells, which are not used in the pulse formation.

They are used as transmission line elements. The buffer

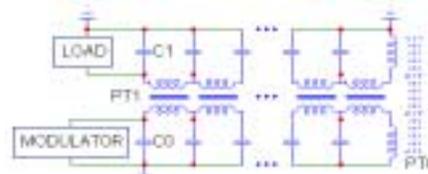
Fig. 1b)

elements always will additionally dissipate some amount of energy.

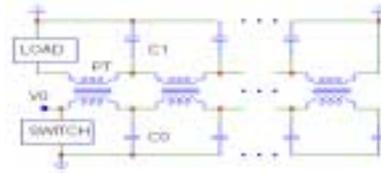
Fig 1c)

Fig. 1: a) scheme of distributed line transformer, b) scheme of a conventional transformer with additional fast cells, and c) scheme of modulator based on coupled pulse forming lines.

Approach presented in Fig 1c) has no buffer elements. One port of primary transmission line can be open. Other port of this line has the switch for discharge of PFN. Here PFN is created by the lumped $C_0$ capacitors and the primary winding of pulse transformers $PT$. The energy is stored in capacitors of the first transmission line. Primary windings of pulse transformers have a floating potential. The charge of the second line is equal to zero. The output pulse is started when the switch is on. The TEM wave starts to propagate toward the end of line. The storage energy is launched partly in the load and partly on the charge of the second line capacitors $C_1$. After the double propagating time, the energy is extracted at the load. This approach was evaluated experimentally as well as by computer simulation. The aim of these experiments was 'proof of principle' i.e. to show a possibility of the pulse formation on the load. The leakage inductors and distributed capacitances here are not parasitic elements. They are parts of pulse forming network. The output waveform of the voltage for two-coupled PFN with a step up voltage transformation is shown in Fig. 2.

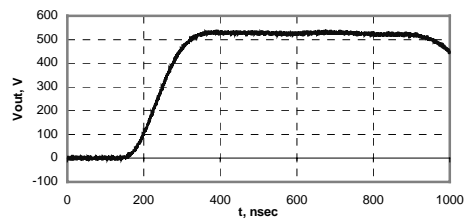

Fig. 2: Output waveform for two-coupled PFN with a step up voltage transformation.

The rise and fall times for this approach can be better as compared to the conventional modulator. The energy transfer efficiency will be higher if a ferrite material is used as a core in this design. The process of the dynamic magnetization of a ferrite material in nanosecond range goes with much lower loss as compared to the iron ribbon.

Evaluation of the total efficiency for a distributed line transformer approach gives a value of ~83% for 1.5 usec pulse duration.

## TRANSFORMERLESS APPROACH

This modulator has no problems, which relate to the use of a ferromagnetic media for transformation of the energy (such as a limitation of the volt-second product, dynamic losses, etc.). Presented approach is based upon a modified Marx method. The modified Marx method charges capacitors in parallel and partially discharges them in series through IGBT assembly [5]. The possible circuit schematics are shown in Fig. 6a) and b).

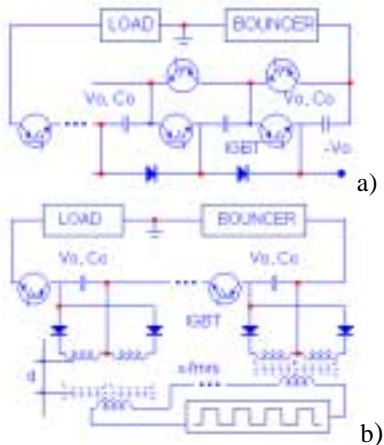

Figure 3: A Marx type modulator concepts.

Because the switch elements are on/off devices, pulse duration is variable, and the efficiency is high. The voltage multiplication of a Marx circuit is equal to the number of cells used. Pulse droop is low depending only on energy storage capacity and load impedance. Compensation for small droop is possible via a bouncer at the low potential side. Extrapolating this transformerless approach for the NLC seems to be attractive especially for time being when the pulse width is not fixed.

The circuit schematic shown in Fig. 3a) is a topology with active recharge switches. They allow a reduction of the charging loss. The topology of Marx type modulator with recharging through isolation transformers is presented in Fig. 3b). Here the NiZn ferrite core transformers are used for charging process. The isolation distance $d$ between primary windings is varied according to the applying pulse voltage.

The main problem of the Marx type modulator is parasitic capacitance. The amplitude of the transient inrush current and its derivative $di/dt$, which flows thought the switches, can destroy solid-state assembly. To get some experience with the transformerless approach a 10 cell solid state Marx type modulator was designed, manufactured and tested [5]. Methods of reduction of inrush current were experimentally evaluated. It was shown that amplitude of transient current can be reduced by factor of ~10 to keep the solid state switch working point within Reverse Biased Safe Operating Area (RBSOA). The volume of solid state Marx type modulators is larger as compared to the transformer concepts due to the absence of ferromagnetic media with a high permeability. The design of multiple beam klystrons is presently under study at SLAC [6]. Authors offer the possibility of considerably lower cathode voltage. The transformerless approach can be effective and low cost solution for multiple beam klystron modulator design.

## CONCLUSION

The discharge energy efficiency can be ~92-93% for the 150 MW and 1.5 μsec output level. However, the energy transfer efficiency is only ~70%. This result was received on Test Bed modulator [2]. For high power modulators the dynamic capacitive effect of the core magnetizing can play a noticeable role. Effect is inherent for all high power pulse transformers. The bandwidth of the PFN-KLYSTRON energy transmission is limited by this effect and by the well-known geometrical effects. Ways of the rise of energy transfer efficiency had been discussed. The use of the inductively coupled distributed lines can overcome the limitation on the pulse transformation bandwidth. Evaluation of total efficiency for a distributed line transformer concept gives a value of >80%. Computer simulations and low voltage experiments suggest the possibility to get a faster rise time on the load with a higher efficiency.

A transformerless modulator approach is presented on a Marx method of the voltage multiplication using on/of IGBT's. Approach has no problems, which relate to the use of a ferromagnetic media for transformation of the energy. The transient process in Marx modulator was analyzed and effective methods of reduction of inrush current was experimentally obtained and kept the solid stage switch working point within of RBSOA.

A new semiconductors based on materials with a wide band gap (for example, Silicon Carbide) will push up the voltage hold-off and make solid state modulator designs at higher voltages simpler and more efficient in the future.

## ACKNOWLEDGEMENTS

This material would not be prepared without the help of R. Koontz, S. Gold, and R. Akre.